# OPTIMIZED IMAGE STEGANALYSIS THROUGH FEATURE SELECTION USING MBEGA

S.Geetha [1] and Dr.N.Kamaraj [2]

[1] Department of Information Technology,
[2] Department of Electrical and Electronics Engineering,
Thiagarajar College of Engineering, Madurai- 625 015, Tamil Nadu, India.
sgeetha@tce.edu, nkeee@tce.edu

## ABSTRACT

*Feature based steganalysis, an emerging branch in information forensics, aims at identifying the presence of a covert communication by employing the statistical features of the cover and stego image as clues/evidences. Due to the large volumes of security audit data as well as complex and dynamic properties of steganogram behaviours, optimizing the performance of steganalysers becomes an important open problem. This paper is focussed at fine tuning the performance of six promising steganalysers in this field, through feature selection. We propose to employ Markov Blanket-Embedded Genetic Algorithm (MBEGA) for stego sensitive feature selection process. In particular, the embedded Markov blanket based memetic operators add or delete features (or genes) from a genetic algorithm (GA) solution so as to quickly improve the solution and fine-tune the search. Empirical results suggest that MBEGA is effective and efficient in eliminating irrelevant and redundant features based on both Markov blanket and predictive power in classifier model. Observations show that the proposed method is superior in terms of number of selected features, classification accuracy and computational cost than their existing counterparts.*

## 1. INTRODUCTION

Steganography is a dynamic tool with a long history and the capability to adapt to new levels of technology. Steganography (covered writing) is the practice of hiding private or sensitive information within something that appears to be nothing out of the usual. Apart from sender and the receiver nobody knows the existence of the message, thereby protecting the data from unauthorized or unwanted viewing. Steganography has evolved into a digital strategy of hiding a file in some form of multimedia, such as an image, an audio file (like a .wav or mp3), and video file or even in TCP header.

Steganography is considered secure if the stego-images do not contain any detectable artifacts due to message embedding. In other words, the set of stego-images should have the same statistical properties as the set of cover-images. If there exists an algorithm that can guess whether or not a given image contains a secret message with a success rate better than random guessing, the steganographic system is considered broken. For a more exact treatment of the concept of steganographic security, the reader is referred to [4][5].

Steganography may provoke negative effects in the outlook of personal privacy, business activity, and national security. The criminals can abuse the technique for planning illegal activities. For example, commercial spies or traitors may thieve confidential trading or technical messages and deliver them to competitors for a great benefit by using hiding techniques. Terrorists may also use related techniques to cooperate for international attacks (like the 911 event in the U.S.) and prevent themselves from being traced. Some others may even think of the possibility of conveying a computer virus or Trojan horse programs via data hiding techniques. Thus, it raises the concerns of enhancing warders' capability and lessening these negative effects by developing the techniques of "steganalysis".

Steganalysis involves detecting the use of steganography inside of a file with little or no knowledge about the steganography algorithm and/or its parameters. Steganographic algorithms sometimes leave a signature in the file that is encoded. With this knowledge presence of secret messages can be detected. It is fair to say that steganalysis is both an art and a science. The art of steganalysis plays a major role in the selection of features or characteristics to test for hidden messages while the science helps in designing the tests themselves.

Many steganalytic techniques which have been developed recently may fall under one of these two classes: ad hoc schemes [20], [21] and feature based schemes that are generic and that use classifiers to differentiate original and stego images [1], [2], [3], [19]. The latter one works in two steps – extraction of generic feature vectors (high pass components, prediction of error...) and training a classifier with these features to separate stego images from original images. Feature based schemes have been studied more recently and found to be more reliable. Since the amount of audit data and the features that such a steganalyser needs to examine is very large, classification by hand is impossible. Analysis is difficult even with computer assistance because extraneous features can make it harder to detect suspicious behavior patterns. Complex relationships exist between the features, which are practically impossible for humans to discover. Some features may contain false correlations, which hinder the process of detecting stego anomalies. Further, some features may be redundant since the information they add is contained in other features. Extra features can increase computation time, and can impact the accuracy of the steganalyser. A steganalyser must therefore reduce the amount of data to be processed. This is extremely important if real-time detection is desired.

Performing feature selection in the context of steganalysis offers several advantages.

– irrelevant features are eliminated; hence a more rational approach can be followed for classifier-based steganalysis:

– The classification performance is enhanced ([9] shows that the addition of irrelevant features decreases the performance of a SVM-based classifier);

–The selected features can help to point out the features that are sensitive to a given steganographic scheme and consequently to bring a highlight on its weaknesses. It can contribute significantly to active steganalysis.

– Computational complexity is greatly reduced for both feature extraction and training the classifier. If we select a set of N features from a set of M, the training time will be M/N (this is due to the linear complexity of classifiers regarding the dimension).The same complexity reduction can also be obtained for feature generation if we assume that the complexity to generate each feature is equivalent.

## 2. FEATURE BASED STEGANALYSERS

This paper aims at demonstrating the feature reduction process to increase the steganalyser's accuracy and reduces the running time, simplifying the classification problem. We show the empirical study on [1], [2], [3] and [19] by feature reduction using MBEGA. We now give a brief recapitulation of these systems below:

### 2.1 WAM Features

The WAM steganalyser in [1] aims to detect the presence of LSB matching payload in a digital image; the method involves computing the residuals from a quasi-Wiener filter: for a two-dimensional signal $S$,

$$R[S] = \frac{\sigma_0^2 S}{\sigma_0^2 + v_R} \tag{1}$$

where $\sigma_0^2$ is the noise variance (for LSB matching affecting every pixel, 0.5), and $v_i$ is a MAP estimate of the local variance of element $i$, based on windows of four different sizes:

$$v_i = \max(0, \min(v_i^3, v_i^5, v_i^7, v_i^9) - \sigma_0^2)$$

$$\text{where } v_i^N = \frac{1}{N^2} \sum_j S_j^2 \tag{2}$$

(for $v_i^N$ i, j is summed over the $N \times N$ neighbourhood of pixel i). The residuals are computed in the wavelet domain: given an input gray-scale image $X$ whose pixel locations are a set $I$, calculate a one-level wavelet decomposition using the 8-tap Daubechies filter. Discard the low frequency sub-band, and denote the horizontal, vertical, and diagonal sub-bands as $H$, $V$ and $D$. Write $R[H]_i$ for the i-th coefficient of the filtered horizontal residual, moments of the coefficients in these filtered subbands: respectively $R[V]_i$ and $R[D]_i$. The WAM features are the absolute central

$$A_m^H = \frac{1}{|I|} \sum_{i \in I} \left| R[H]_i - \ddot{R}[\ddot{H}] \right|^m,$$

$$A_m^V = \frac{1}{|I|} \sum_{i \in I} \left| R[V]_i - \ddot{R}[\ddot{V}] \right|^m,$$

$$A_m^D = \frac{1}{|I|} \sum_{i \in I} \left| R[D]_i - \ddot{R}[\ddot{D}] \right|^m \tag{3}$$

The first nine moments in each sub-band, form the 27-dimensional feature vectors used in [1].

In [1] a Fisher Linear Discriminator (FLD) is trained on the features of some cover and stego images, to make a detector for the presence of LSB matching steganography. Extremely sensitive detection is reported, with accuracy of around 90% when the LSB matching payload is 25% of the maximum (0.25 bits per pixel), and near-perfect detection with payloads at 50%.

## 2.2 IQM Features

As for the selection of quality measures, the authors of [2] used several (26 in total) metrics for investigation to predict compression, blur and noise artifacts. From these measures the authors experimented out the ones that served well the purpose of steganalysis. The image quality metrics (IQMs) that were employed are listed in Table 1.

In the design phase of the steganalyzer, the authors have regressed the normalized IQM scores to, respectively, -1 and 1, depending upon whether an image did not or did contain a message. Similarly, IQM scores were calculated between the original/stego images and their filtered versions. An average detection rate of 77% has been reported by this steganalyser.

## 2.3 Fridrich's Features

The features proposed by Fridrich et al [3] are computed as follows: a vector functional F is applied to the stego JPEG image J1 and to the virtual clean JPEG image J2 obtained by cropping J1 with a translation of 4 × 4 pixels. The feature is finally computed taking the L1 of the difference of the two functionals:

$$f = \|F(J_1) - F(J_2)\|_{L1} \qquad (4)$$

The functionals used in this paper are described in the Table 2. This model also used SVM classifier and an average detection rate of 88% has been observed.

Table 1. List of 19 IQMs used as features for steganalysis in [2]

| Image Quality Measures used for Steganalysis in [2] ||
|---|---|
| Minkowsky Metric $\gamma = 2$ <br> Minkowsky Metric 1 <br> Maximum Difference <br> Sorted Maximum Difference <br> Czenakoski <br> Structural Content <br> Cross Correlation <br> Image Fidelity <br> Angle Mean <br> Angle Standard Deviation <br> Spectral Magnitude <br> Spectral Phase | Weighted Spectral Distance <br> Median Block Spectral Magnitude <br> Median Block Spectral Phase <br> Median Block Weighted Spectral Distance <br> Normalized Absolute Error (HVS) <br> Normalized Mean Square Error (HVS) <br> HVS Based L2 |

In this paper we propose to introduce the feature selection phase into these steganalysers as a pre-processing phase and thereby optimize their performances.

### 2.3    Higher order Statistics Features

The work in [19] decomposes the image based on separable quadrature mirror filters (QMFs). This decomposition splits the frequency space into multiple scales and orientations. This is accomplished by applying separable low-pass and high-pass filters along the image axes generating a vertical, horizontal, diagonal and low-pass sub-band. Subsequent scales are created by recursively filtering the low-pass sub-band. The vertical, horizontal, and diagonal sub-bands at scale $i = 1,...,n$ are denoted as $V_i(x, y), H_i(x, y)$ and $D_i(x, y)$ respectively.

Given this image decomposition, the statistical model is composed of the mean, variance, skewness and kurtosis of the sub-band coefficients at each orientation and at scales $i = 1,...,n-1$. These statistics characterize the basic coefficient distributions. The second set of statistics is based on the errors in an optimal linear predictor of coefficient magnitude. Then the sub-band coefficients are correlated to their spatial, orientation and scale neighbors.

The vertical band being $V_i(x, y)$, at scale $i$, a linear predictor for the magnitude of these coefficients in a subset of all possible neighbors is given by:

$$V_i(x, y) = w_1 V_i(x-1, y) + w_2 V_i(x+1, y)$$
$$+ w_3 V_i(x, y-1) + w_4 V_i(x, y+1)$$
$$+ w_5 V_{i+1}(x/2, y/2) + w_6 D_i(x, y)$$
$$+ w_7 D_{i+1}(x/2, y/2)$$

(5)

where $w_k$ denotes scalar weighting values. This linear relationship is expressed more compactly in matrix form as:

$$V = Qw \tag{6}$$

where the column vector $w = (w_1...w_7)^T$, the vector $V$ contains the coefficient magnitudes of $V_i(x, y)$ strung out into a column vector, and the columns of the matrix $Q$ contain the neighbouring coefficient magnitudes as specified in Equation (5) also strung out into column vectors. The coefficients are determined by minimizing the quadratic error function:

$$E(w) = [V - Qw]^2 \tag{7}$$

This error function is minimized by differentiating with respect to $w$:

$$dE(w)/dw = 2Q^T[V - Qw] \tag{8}$$

setting the result equal to zero, and solving for $w$ to yield:

$$w = (Q^T Q)^{-1} Q^T V. \tag{9}$$

The log error in the linear predictor is then given by:

$$E = \log_2(V) - \log_2(|Qw|) \tag{10}$$

It is from this error that additional statistics are collected namely the mean, variance, skewness, and kurtosis. This process is repeated for each vertical sub-band at scales $i = 1,...,n-1$, where at each scale a new linear predictor is estimated. A similar process is repeated for the horizontal and diagonal sub-bands. The linear predictor for the horizontal sub-bands is of the form:

$$H_i(x, y) = w_1 H_i(x-1, y) + w_2 H_i(x+1, y)$$
$$+ w_3 H_i(x, y-1) + w_4 H_i(x, y+1)$$
$$+ w_5 H_{i+1}(x/2, y/2) + w D_i(x, y)$$
$$+ w_7 D_{i+1}(x/2, y/2)$$

(11)

and for the diagonal subbands:

$$D_i(x, y) = w_1 D_i(x-1, y) + w_2 D_i(x+1, y)$$
$$+ w_3 D_i(x, y-1) + w_4 D_i(x, y+1)$$
$$+ w_5 D_{i+1}(x/2, y/2) + w H_i(x, y)$$
$$+ w_7 V_{i+1}(x/2, y/2)$$

(12)

The same error metric, Equation (10), and error statistics computed for the vertical sub-ands, are computed for the horizontal and diagonal bands, for a total of $12(n-1)$ error statistics. Combining these statistics with the $12(n-1)$) coefficient statistics yields a total of $24(n-1)$ statistics that form a feature vector which is used to discriminate between images that contain hidden messages and those that do not. Experiments were done with the value of n=4, i.e, wavelet decomposition of the image was done to level 4. Totally 72 statistical features were employed for steganalysis.

A non-linear Support Vector Machine classifier with 1.0% false positive rate produced good results with an average detection rate of 80% for JSteg, OutGuess and a moderate detection rate for EzStego as 55% and poor results for LSB systems i.e. only 62%. With less payload capacity, the detection rate was not promising and was low around 5-10% only.

## 3. FEATURE SELECTION

Generally, a typical feature selection method consists of four components: a subset generation or search procedure, an evaluation function, a stopping criterion, and a validation procedure. This general process of feature selection is illustrated in Figure 1. A key issue for feature selection algorithm is how to search the space of feature subsets which is exponential in the number of features.

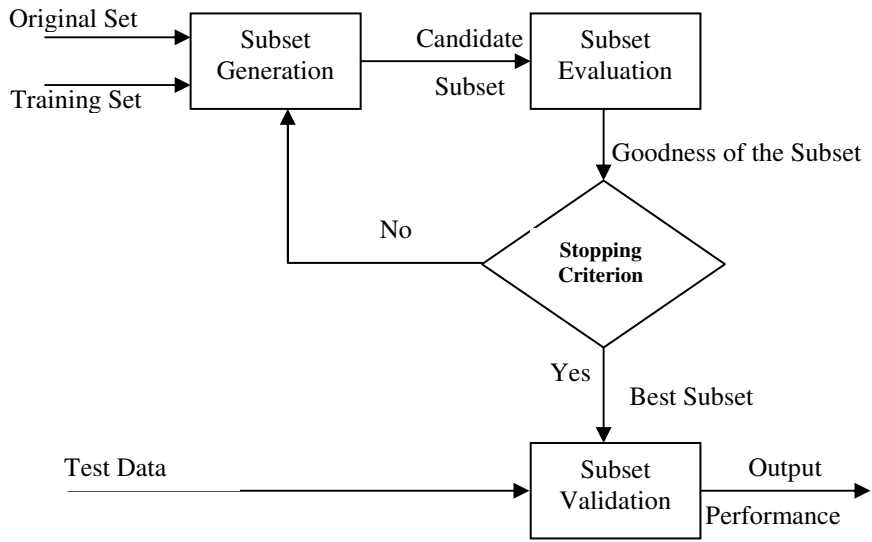

Figure 1. General Procedure for Feature Selection Process

Genetic Algorithm (GA)[7] one of the commonly used modern stochastic global search techniques, has well known ability to produce high quality solution within tractable time even on complex problems. It has been naturally used for feature and has shown promising performance [8][9]. Unfortunately, due to the inherent nature of GA, it often takes a long time to locate the local optimum in a region of convergence and may sometimes not find the optimum with sufficient precision. One way to solve this problem is to hybridize GA with some memetic operations (also known as local search operations) [10] which are capable of fine-tuning and improving the solutions generated by the GA to make them more accurate and efficient. Particularly, they not only converge to high quality solutions, but also search more efficiently than their conventional counterparts [10].

In this work, we propose the use of a novel Markov blanket embedded GA (MBEGA) feature selection algorithm [11] for steganalysis problem. MBEGA uses Markov blanket to fine-tune the search by adding the relevant features or removing the redundant and/or irrelevant features in the GA solutions. This memetic algorithm takes advantage of both Markov blanket and GA wrapper feature selection with the aim to improve classification performance and accelerate the search to retain relevant and remove redundant features.

Listing 1. Markov Blanket Embedded Genetic Algorithm (MBEGA) for Gene Selection

***Markov Blanket Embedded Genetic Algorithm (MBEGA)***
**BEGIN**
  (1) **Initialize:** *Randomly generate an initial population of feature subsets encoded with binary string.*
  (2) **While***(not converged or computational budget is not exhausted)*
  (3) *Evaluate fitness of all feature subsets in the population based on $J(S_c)$.*
  (4) *Select the elite chromosome $c_b$ to undergo Markov blanket based memetic operation.*
  (5) *Replace $c_b$ with improved new chromosome $C_b^n$ using Lamarckian learning.*
  (6) *Perform evolutionary operators based on restrictive selection, crossover, and mutation.*
  (7) **End While**
**END**

Listing 2. ADD and DEL memetic operators employed for Gene Selection

*Add Operator:*
**BEGIN**
  (1) *Rank the features in Y in a descending order based on C-correlation value.*
  (2) *Select a feature $Y_i$ in Y using linear ranking selection [38] so that the larger the C-correlation of a feature in Y the more likely it will be selected.*
  (3) *Add $Y_i$ to X.*
**END**

*Del Operator:*
**BEGIN**
  (1) *Rank the features in X in a descending order based on C-correlation value.*
  (2) *Select a feature $X_i$ in X using linear ranking selection [38] so that the larger the C-correlation of a feature in X the more likely it will be selected.*
  (3) *Eliminate all features in $X-\{X_i\}$ which are in the approximate Markov blanket of $X_i$. If no feature is eliminated, remove $X_i$ itself.*
**END**

Listing 3. Complete Markov Blanket Based Memetic Operation

***Markov Blanket Based Memetic Operation***
**BEGIN**
  (1) Select the elite chromosome $C_b$ to undergo memetic operations.
  (2) **For** $l = 1$ to $L^2$
  (3) Generate a unique random pair $\{a,d\}$ where $0 < a, d < L$.
  (4) Apply $a$ times Add on $C_b$ to generate a new chromosome $C_b^{'}$.
  (5) Apply $d$ times Del on $C_b^{'}$ to generate a new chromosome $C_b^n$.
  (6) Calculate fitness of modified chromosome $C_b^n$ based on $J(S_c)$.
  (7) **If** $C_b^n$ is better than $C_b$ either on fitness or number of features
  (8) Replace the genotype $C_b$ with $C_b^n$ and stop memetic operation.
    **End If**
  **End For**
**END**

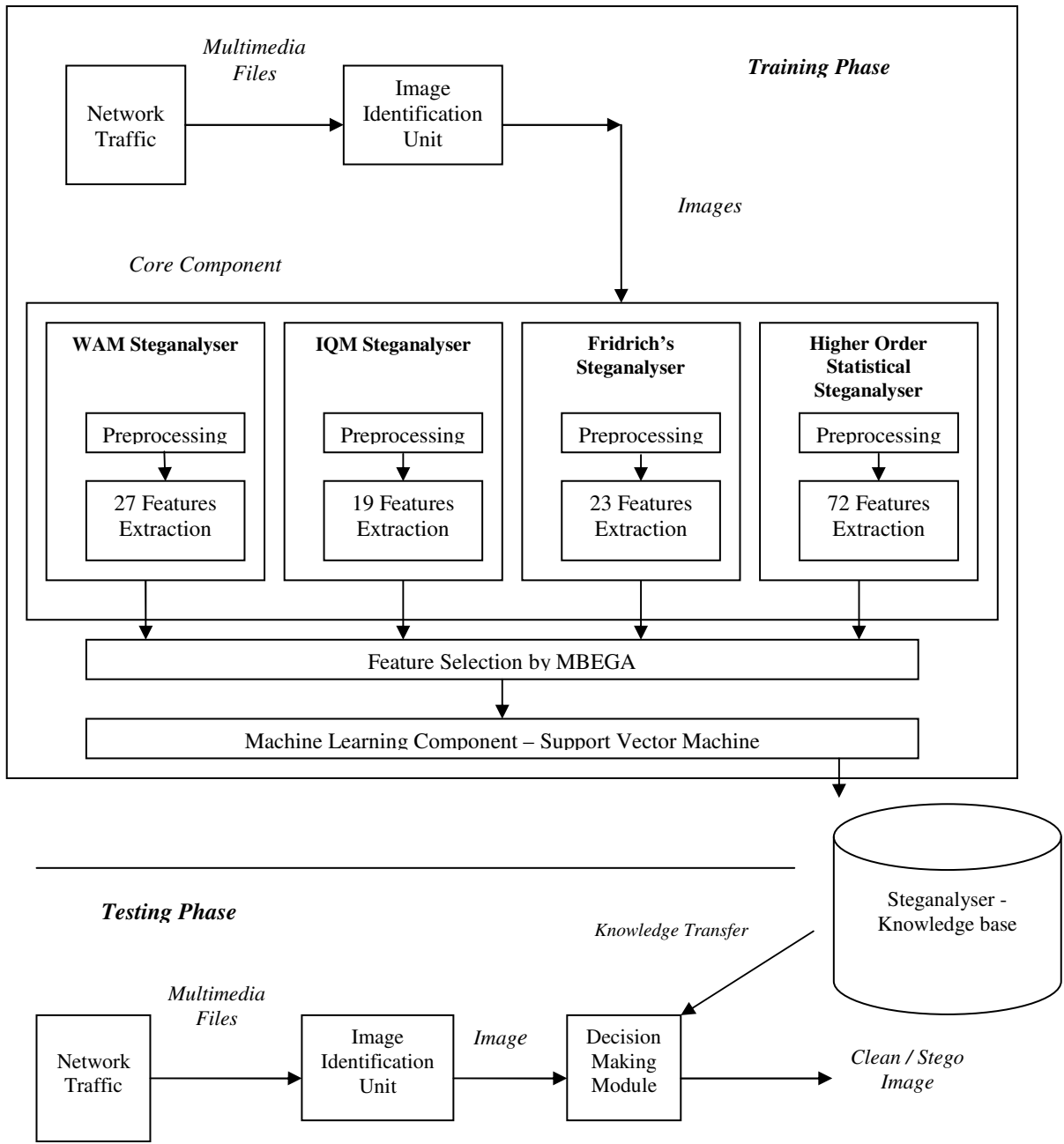

Figure 2. Architecture of the proposed system with MBEGA Feature Selection

## 4. SYSTEM AND METHODOLOGY

In this section we present an overview about the proposed system architecture and the MBEGA algorithm employed for steganalysis. The pseudo code of the proposed memetic algorithm, the Markov Blanket-Embedded GA (MBEGA). is outlined in Listing 1, 2 and 3.

At the start of the MBEGA search, an initial GA population is initialized randomly with each chromosome encoding a candidate feature subset. Using binary encoding, a bit of '1' ('0') implies the corresponding feature is selected (excluded). The fitness of each chromosome is then obtained using an objective function based on the induction algorithm:

Fitness(c) = J(Sc)                                                                                                         (13)

where Sc denotes the selected feature subset encoded in a chromosome c, and the feature selection objective function J(Sc) evaluates the significance for the given feature subset Sc. Here, J(Sc) is the generalization error obtained for Sc which can be estimated using cross validation or bootstrapping techniques. When two chromosomes are found having similar fitness, the one with a smaller number of selected features is given higher chance of surviving to the next generation.

In each GA generation, the elite chromosome, i.e., the one with the best fitness value then undergoes Markov blanket based memetic operators/local search in the spirit of Lamarckian learning [12]. The Lamarckian learning forces the genotype to reflect the result of improvement through placing the locally improved individual back into the population to compete for reproductive opportunities. Two memetic operators, namely an Add operator that inserts a feature into the elite chromosome, and a Del operator that removes existing features from the elite chromosome, are introduced in the MBEGA.

## 5. EMPIRICAL RESULTS

### 5.1    Data Source:

To evaluate the performance of the proposed method, image samples from [13] were used; a large database of 250 images with categories like Animals, Birds, Buildings, Nature (Sky and cloud, Flowers, Fruits and Face is used. The data set comprised of images with diverse nature having various degrees of texture, color, brightness and intensity. This database is augmented with the stego versions of these images using five popular data embedding systems, i.e., three watermarking and two steganographic techniques with a 50% payload (0.5bpp). The watermarking techniques are Cox [14], Digimarc [15] and PGS [16]. The steganographic methods are StegHide [17] and S-Tools [18]. The rationale of using these tools was their popularity, high embedding capacity, free availability, wide usage, and applicability to images.

### 5.2    Training and Testing Data

In order to conduct an experimental setting, different sets of 250 images were used. By embedding separately with each watermarking method, we got an overall of 1500 marked records. For each individual method, a mixture containing 150 embedded records and 150 original records are used for training, while an independent mixture containing 100 original and 100 embedded records is kept for testing purposes.

### 5.3    Results and Discussions

The model is incorporated in Java JGAP and the algorithm is implemented as per the framework proposed. SVM classifier engine is employed to build the knowledge base, as prescribed in the actual works [3], [19] owing to its superior performance than other soft computing paradigms. Table 3 provides the amount of feature reduction achieved in each steganalytic scheme. The reduction in computational complexity i.e. running time is also shown in Table 3. The performance comparisons are shown in Table 4. The proposed scheme, the feature selection based on MBEGA, achieves very satisfactory results, all in terms of feature reduction, classification accuracy and computational cost. It could be observed that on an average 55.897% of feature reduction is achieved. This contributes to the significant decrease in computational cost and also in the simplification of the solution space. The running time consumed during training phase is cut down by 54.42%, which is a substantial improvement over the existing systems. The detection accuracy (Table 4) is almost close to the existing systems even with the proposed system – WAM and Fridrich's method has no compromise over the detection power while IQM method shows an increase of about 3% from 77% to 80%, and higher order statistics method shows an increase by 3.3%, i.e., from 81.07% to 84.4%, which is promising. Figure 2 shows the number of features before and after feature selection. Figure 3 depicts the classification accuracy before and after MBEGA feature selection. This reveals the fact that the feature reduction through MBEGA introduced in our system greatly enhances the performance in terms of computational complexity and detection accuracy.

## 6. CONCLUSION

This paper has presented the use of MBEGA for dimensionality reduction by feature selection in the framework of steganalysis. The strength of the system has been demonstrated on up to 1750 images. The major findings of the proposed work may be summarized as follows:

A set of 27 WAM based features, 19 IQM based features, 23 Fridrich's method features and 72 higher order statistics as features has been subjected to dimensionality reduction step, yielding a subset of stego sensitive features, a set of 12, 6 10 and 41 features only respectively. An average reduction of 55.89% has been reported.
The computational complexity is minimized by 54.42% over the existing systems.
The detection accuracy shows no significant compromise over the existing systems. IQM method and higher order statistical methods report an increase by 3% in classifier accuracy.

The future direction of this research may be experimenting with the usage of other feature selection algorithms on various other powerful steganalysers. Audio systems may also be targeted for optimization. An analysis of the reduced feature sets could help design a more generic steganalysis method using a "low" number of features.

Table 2. List of 23 features used for steganalysis in [3]

| Functional/ Feature name | Functional F |
|---|---|
| Global Histogram | $H/\|H\|$ |
| Individual histogram for 5 DCT Modes | $h^{21}/\|h^{21}\|, h^{12}/\|h^{12}\|, h^{13}/\|h^{13}\|, h^{22}/\|h^{22}\|, h^{31}/\|h^{31}\|$ |
| Dual histogram for 11 DCT values (-5,…,5) | $g^{-5}/\|g^{-5}\|, g^{-4}/\|g^{-4}\|, g^{-3}/\|g^{-3}\|, g^{-2}/\|g^{-2}\|, g^{-1}/\|g^{-1}\|,$ $g^{0}/\|g^{0}\|, g^{1}/\|g^{1}\|, g^{2}/\|g^{2}\|, g^{3}/\|g^{3}\|, g^{4}/\|g^{4}\|, g^{5}/\|g^{5}\|$ |
| Variation | $V$ |
| L1 and L2 blockiness | $B_1, B_2$ |
| Co-occurence | $N_{00}, N_{01}, N_{11}$ |

Table 3. Running Time and Feature Reduction Comparison - Actual systems vs. Proposed system

| Steganalytic Scheme | Original feature set – without MBEGA | Selected feature set- with MBEGA | % Reduction in Features | Running time in the original system (ms) | Running time in the proposed system (ms) | % Reduction in time |
|---|---|---|---|---|---|---|
| *WAM method* | 27 features | 12 features | 55.56% | 980 | 400 | 59.18% |
| *IQM method* | 19 features | 6 features | 68.42% | 170 | 72 | 57.64% |
| *Fridrich's method* | 23 features | 10 features | 56.52% | 860 | 385 | 55.23% |
| *Higher Order Statistics method* | 72 features | 41 features | 43.06% | 1830 | 995 | 45.63% |

Table 4. Detection accuracy comparison of the actual systems vs. proposed image steganalyzer model at 50% payload

| Data Hiding Technique | *WAM method* | | *IQM method* | | *Fridrich's method* | | *Higher Order Statistics method* | |
|---|---|---|---|---|---|---|---|---|
| | Without MBEGA | With MBEGA | Without MBEGA | Without MBEGA | With MBEGA | With MBEGA | Without MBEGA | With MBEGA |
| **Cox** | 86% | 86.2% | 70% | 72.73% | 71.81% | 73.66% | 74.51% | 79.5% |
| **DigiMarc** | 89% | 87.5% | 80% | 85.21% | 85% | 83.66% | 77.2% | 86.5% |
| **PGS** | 94% | 94.3% | 85% | 91.66% | 92% | 86.5% | 85.8% | 88.9% |
| **StegHide** | 95% | 95.6% | 70% | 93.25% | 94% | 72.5% | 76% | 77.2% |
| **S-Tools** | 96% | 96% | 75% | 90.21% | 92% | 81.33% | 77% | 78% |
| **Clean** | 93% | 94% | 80% | 92.33% | 93% | 83.33% | 95% | 96.3% |
| **Average accuracy** | 92% | 92% | 77% | 88% | 88% | 80% | 81.07% | 84.4% |
| **% increase/decrease in accuracy** | No change | | 3% increase | | No change | | 3.3% increase | |

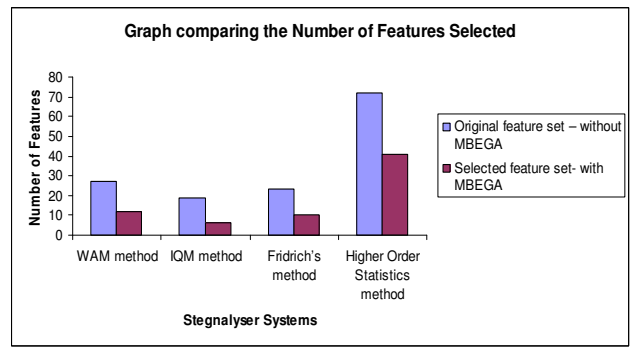

Figure 2. Comparison of the number of features – with and with out MBEGA

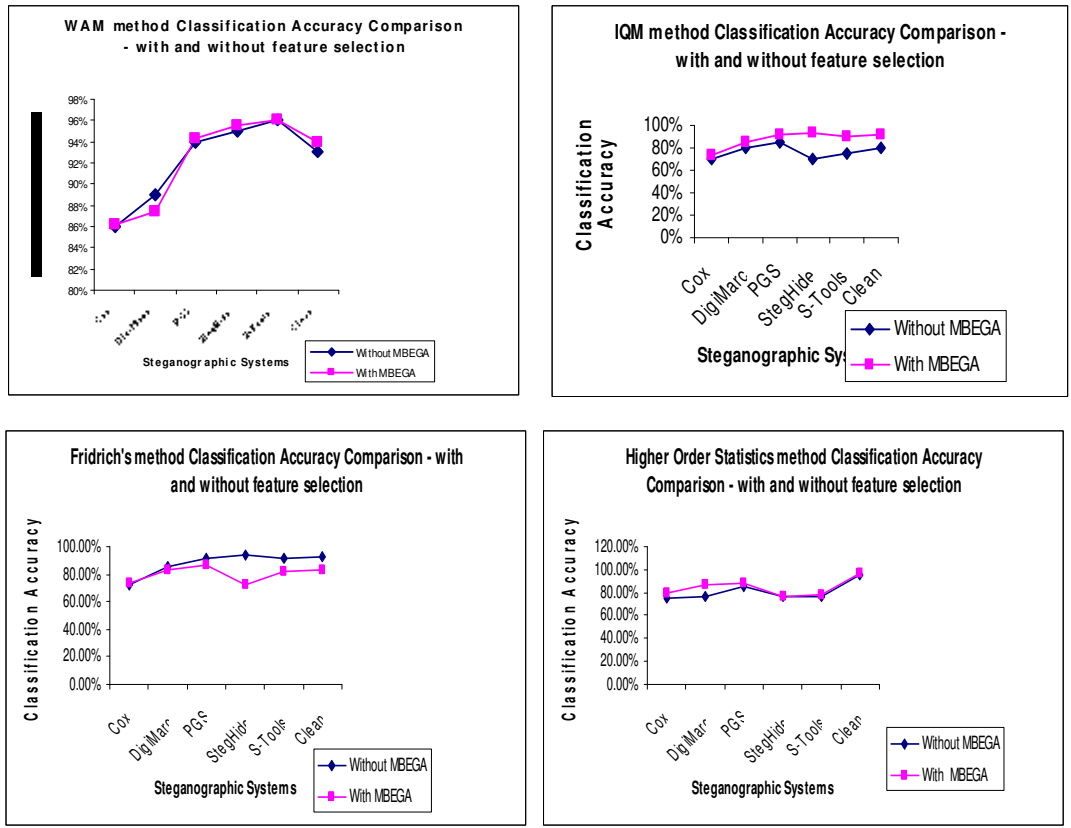

Figure 3. Comparison of the classification accuracy of the various steganalysers – with and with out MBEGA

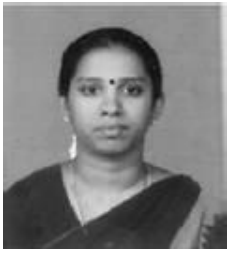S.Geetha received the B.E., and M.E., degrees in Computer Science and Engineering in 2000 and 2004, respectively, from the Madurai Kamaraj University and Anna University of Chennai, India. In July 2004, she joined the Department of Information Technology at Thiagarajar College of Engineering, Madurai, India. She has published more than 20 papers in reputed IEEE International Conferences and refereed Journals. She joins the review committee for IEEE Transactions on Information Forensics and Security, IEEE Transactions on Image Processing, Taylor and Francis – Information Security, Springer Verlag – Multimedia Tools and Applications, Elsevier Information Sciences etc. She was an editor for the ICCIIS 2007 – 1st Indian Conference on Computational Intelligence and Information Security. Her research interests include multimedia security, intrusion detection systems, machine learning paradigms and information forensics. She is a recipient of University Rank and Academic Topper Award in B.E. and M.E. in 2000 and 2004 respectively.

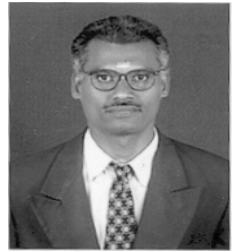Dr. N.KAMARAJ is an Associate Professor in Electrical and Engineering Department , Thiagarajar College of Engineering, Madurai, Tamilnadu, India. He obtained B.E. degree in Electrical and Electronics Engineering and M.E. degree in Power System Engineering from Madurai Kamaraj University in the year 1988 and 1994 respectively. He obtained Ph.D. Degree in the Power System Security Assessment in the year 2003 from Madurai Kamaraj University. Currently he is heading the department of Electrical and Electronics Engineering in Thiagarajar College of Engineering. His research areas include Security Assessment using Neural Network, Fuzzy logic and Genetic Algorithm. He has published 11 papers in the International & National journal and presented 10 papers in the International conferences and 22 papers in the National conferences. He is the recipient of Merit award from IEEE – Computer Society for CSIDC 2003 as best advisor for the team contested in CSIDC. Also he has received Gold medal and Corps subject award from Institution of Engineers (India) for the year 2003.